\documentclass[journal]{IEEEtran}

\IEEEoverridecommandlockouts 
\usepackage{graphicx,latexsym,subfigure,epsfig,latexsym,subfigure,amsmath,cite,amssymb}
\usepackage{amsthm,mathrsfs}
\usepackage{algorithmic}
\usepackage{graphicx}
\usepackage{epstopdf}
\usepackage{color}
\usepackage{CJK}
\usepackage[ruled,linesnumbered]{algorithm2e}
\usepackage{mathrsfs}
\usepackage{bm}
\usepackage{float}
\usepackage{setspace}
\usepackage{color}
\usepackage{subfigure}
\usepackage{balance}
\usepackage{subeqnarray}
\usepackage{cases}
\usepackage{hyperref}

\begin{document}
\title{UAV-Aided Cellular Communications with Deep Reinforcement Learning Against Jamming}
\author{\IEEEauthorblockN{\footnotesize Xiaozhen Lu\IEEEauthorrefmark{1}, Liang Xiao\IEEEauthorrefmark{1}, Canhuang Dai\IEEEauthorrefmark{1},
Huaiyu Dai\IEEEauthorrefmark{2}\\
\IEEEauthorblockA{\IEEEauthorrefmark{1}\footnotesize Dept. of Information and Communication Engineering, Xiamen University, Xiamen, China. Email: lxiao@xmu.edu.cn}\\
\IEEEauthorblockA{\IEEEauthorrefmark{2}Dept. of Electrical and Computer Engineering, NC state University, Raleigh, NC. Email: huaiyu\_dai@ncsu.edu}
}
}
\maketitle
\begin{abstract}
Cellular systems are vulnerable to jamming attacks, especially smart jammers that choose their jamming policies such as the jamming channel frequencies and power based on the ongoing communication policies and network states. In this article, we present an unmanned aerial vehicle (UAV) aided cellular communication framework against jamming. In this scheme, UAVs use reinforcement learning methods to choose the relay policy for mobile users in cellular systems, if the serving base station is heavily jammed. More specifically, we propose a deep reinforcement learning based UAV relay scheme to help cellular systems resist smart jamming without being aware of the jamming model and the network model in the dynamic game based on the previous anti-jamming relay experiences and the observed current network status. This scheme can achieve the optimal performance after enough interactions with the jammer. Simulation results show that this scheme can reduce the bit error rate of the messages and save energy for the cellular system compared with the existing scheme.
\end{abstract}


\IEEEpeerreviewmaketitle
\section{Introduction}
\begin{figure*}[!htbp]
\begin{center}
\includegraphics[height=2in]{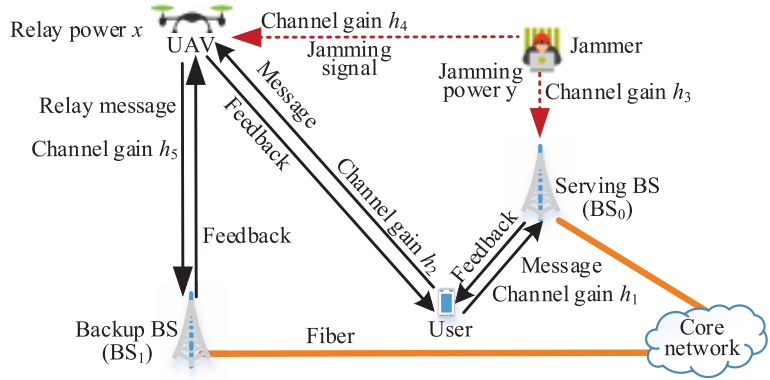}\\
\caption{UAV-aided cellular systems against jamming attacks.}\label{fig:model}
\end{center}
\end{figure*}
Cellular systems such as the 5th generation (5G) and 4th generation (4G) have to support booming computation-intensive applications such as augmented reality (AR) games and are vulnerable to jamming attacks due to the high user mobility in large-scale dynamic cellular networks \cite{Zeng2016Wireless}. Jammers send faked or replayed jamming signals to block the ongoing communications, exhaust the battery levels of mobile users, threaten user privacy, and further perform other attacks such as man-in-the-middle attacks. Jammers can be static or reactive. In particular, smart jammers as an advanced and most dangerous type of reactive jammers use smart radio devices such as universal software radio peripheral (USRP) and machine learning techniques to infer the cellular system defense policy and then attack it accordingly \cite{Huo2017Jamming}.

Unmanned aerial vehicles (UAVs) can help cellular systems resist jamming due to the high altitude, mobility, and line-of-sight (LOS) channels to the mobile users \cite{Dixon2012Optimizing, Zeng2016Throughput, Feng2017Positioning, Lu2018UAV}. More specifically, UAVs relay the messages of mobile users if the serving base stations (BSs) are seriously blocked. In the UAV-aided cellular system, a UAV has to choose its relay policy such as the relay power without being aware of the jamming model and radio channel states. However, existing communication optimization schemes such as the convex optimization based UAV relay scheme as presented in \cite{Dixon2012Optimizing} requires the relay UAV to accurately know the speeds and positions of all the UAVs and the UAV channel model, and their communication performance degrades in dynamic UAV networks.
This issue can be addressed by reinforcement learning (RL) techniques such as Q-learning and policy hill climbing (PHC) \cite{Lv2017Anti,Lu2018UAV}, as the repeated UAV relay process in the dynamic cellular system against jamming can be formulated as a Markov decision process (MDP) \cite{Xie2018User}.

Reinforcement learning techniques have been applied in the anti-jamming relay selection and power control in wireless networks \cite{Dams2016Jamming, Wu2012Anti} and vehicular ad hoc networks (VANETs) \cite{Lu2018UAV}. For instance, in the RL based UAV-aided VANET system as proposed in \cite{Lu2018UAV}, the UAV applies the PHC algorithm to choose whether to relay a message for an onboard unit in the VANET if the area is seriously jammed or interfered. This scheme can improve the signal-to-interference-plus-noise-ratio (SINR) of the onboard unit signals and reduce the bit error rate (BER) of the messages. However, this scheme will suffer from a long learning time and its performance will be degraded in cellular systems, due to the random exploration at the beginning, the estimation error and delay regarding the network state, and the reward in the dynamic game.

The user-UAV link and the BS-UAV link usually have better channel states due to the LOS propagation of the UAV, compared with the link of the user and the serving BS at a fixed location that is severely blocked by jamming attacks \cite{Lu2018UAV}. In this article, we propose a deep reinforcement learning based UAV relay scheme to choose its relay power against jamming attacks, including smart jamming. This scheme uses deep Q-network (DQN) \cite{He2017Deep} and transfer learning \cite{Lu2018UAV} to accelerate the learning speed of the PHC based anti-jamming UAV relay algorithm named HPUR in \cite{Lu2018UAV}. More specifically, the UAV exploits the deep convolutional neural network (CNN) to compress the high-dimensional state space, applies the experience replay technique to update the CNN parameters, uses transfer learning to initialize the CNN weights with the previous anti-jamming relay experiences in similar scenarios.

In the deep RL based relay scheme, the UAV formulates the current state with the BER of the message received by the serving BS, that sent from the user to the UAV, that for the message from the UAV to the backup BS, the estimated channel power gains and the estimated jamming power. The UAV chooses the relay power based on the current state and the Q-values for each relay policy, which is the output of the CNN. This scheme enables a UAV to decide the optimal relay power without knowing the jamming model and the network model in a dynamic anti-jamming relay game.

The deep RL based relay scheme can optimize the relay power via error-and-trials against static jammers, reactive jammers and smart jammers that apply RL to choose the jamming power with the goal to minimize the UAV utility with less jamming costs. We provide the BER performance bound of the user messages and discuss the computation complexity. Simulation results show that this relay scheme can efficiently reduce the BER and save energy for cellular systems to resist smart jamming attacks compared with the benchmark relay scheme in \cite{Lu2018UAV}.

The contributions of this work can be summarized as follows:
\begin{itemize}
  \item We design a deep RL based UAV relay scheme that uses deep reinforcement learning and transfer learning to optimize the relay power against jamming without requiring the knowledge of the network status and the jamming model.
  \item We prove that the relay scheme can achieve the optimal relay power via error-and-trials, and provide the BER performance bound. We also evaluate the computation complexity of the proposed relay scheme.
\end{itemize}

This article is organized as follows. We review the related work and present the UAV-aided cellular system. We then propose a deep RL based relay scheme for cellular systems and evaluate its performance. Finally, we conclude this article and identify the future work.
\begin{table*}[!htbp]
  \caption{Summary of the learning-based anti-jamming wireless communication methods}
\newcommand{\tabincell}[2]{\begin{tabular}{@{}#1@{}}#2\end{tabular}}
  \centering
  \begin{tabular}{|l|l|l|l|l|}\hline
Learning techniques & Action & Performance & Application & Ref\\\hline
Q-learning & \tabincell{l}{Power allocation\\Relay policy\\Channel selection\\Duty cycle selection} &\tabincell{l}{BER\\SINR\\Capacity\\Safe rate\\Secrecy capacity\\Percentage of payoff}&\tabincell{l}{VANETs\\UAV systems\\Cellular systems\\Cognitive radio networks}& \tabincell{l}{ \cite{Lv2017Anti,Xie2018User,Lu2018UAV}\\ \cite{Wu2012Anti,Athukoralage2016Regret}}\\\hline
\tabincell{l}{PHC/WoLF-PHC} & \tabincell{l}{Relay policy\\Power allocation} & \tabincell{l}{BER\\SINR\\Safe rate\\Secrecy capacity} & \tabincell{l}{VANETs\\UAV systems}& \cite{Xie2018User,Lu2018UAV}\\\hline
Randomized WMA & \tabincell{l}{Transmit policy} & \tabincell{l}{Successful rate} & Wireless networks & \cite{Dams2016Jamming}\\\hline
DQN & \tabincell{l}{User selection\\Power allocation\\Resource allocation} & \tabincell{l}{SINR\\Sum rate\\Safe rate\\Secrecy capacity\\Energy efficiency} & \tabincell{l}{UAV systems\\Wireless networks} & \cite{He2017Deep,Xie2018User}\\\hline
Bayesian & \tabincell{l}{Monitor policy\\Node classification} & \tabincell{l}{Accuracy rate\\Communication overhead} & \tabincell{l}{VANETs} & \cite{Sedjelmaci2017Intrusion}\\\hline
\end{tabular}\label{tabb}
\end{table*}
\section{Related work}
UAVs can relay messages for ground terminals against jamming attacks in cellular systems. For instance, the mobile relaying systems as proposed in \cite{Zeng2016Throughput} optimize the UAVs transmit power and trajectory to resist jamming and improve the throughput of cellular systems. The 5G communication scheme in \cite{Feng2017Positioning} uses UAVs to relay the messages for mobile devices to resist jamming and achieve lower time complexity. The UAV-aided VANETs in \cite{Lu2018UAV} presents a hotbooting PHC based relay algorithm and uses UAV to relay the messages for onboard units to resist smart jamming, which can reduce the BER of the messages and thus improves the SINR of the VANET.

UAVs can use reinforcement learning techniques to resist jamming attacks. For instance, the UAV relay scheme in \cite{Athukoralage2016Regret} uses the regret-based Q-learning algorithm to choose the transmission duty cycle to relay the messages and increase the communication capacity for cellular systems against jamming attacks. A cache-enabled communication scheme as developed in \cite{He2017Deep} applies DQN for interference alignment and user selection to provide user cooperation and resist jamming. The anti-jamming scheme as designed in \cite{Lv2017Anti} applies Q-learning to select the transmit power to resist jamming attacks and achieve higher SINR of the UAV systems. A UAV communication system as developed in \cite{Xie2018User} uses DQN to choose the transmit power against subjective smart attackers to improve the safe rate and the secrecy capacity.

UAVs can also apply supervised learning techniques such as randomized weighted majority algorithm (WMA) and Bayesian to prevent wireless networks from jamming attacks. For example, the UAV relay scheme as presented in \cite{Dams2016Jamming} uses RWMA to decide whether to relay the messages and improve the successful transmission rate for cellular systems against jamming attacks. The UAV relay assisted VANET system in \cite{Sedjelmaci2017Intrusion} uses a nonparametric Bayesian method to resist jamming attacks and save the UAV energy consumption. However, these schemes will suffer from computation and communication costs due to a large amount of training data and complicated features extraction process, as summarized in Table \ref{tabb}.

\section{System model}
\subsection{Network Model}
As shown in Fig. \ref{fig:model}, a user such as a smart-phone with limited caching resources and battery life has to send real-time messages such as videos and AR game information to the server in the core network. Each BS at the fixed location is connected via fibers to each other and the core network. The current serving BS of the user is denoted by BS$_0$. A UAV monitors the status of the BSs in the area and helps relay the user messages if a BS is seriously blocked by a jammer. The UAV moves to a position that is farther away from the jammers compared with BS$_0$ and chooses the relay power to relay the message to a backup BS denoted by BS$_1$, which is assumed to be far away from the jammer and can receive the relay message from the UAV.

The user sends a message with transmit power $P$ at time slot $k$ to the serving BS connecting to the server in the core network via fibers. Upon decoding the user message from BS$_0$, the server measures its BER $\rho_1^{(k)}$ and sends $\rho_1^{(k)}$ to the UAV via BS$_1$.

The UAV chooses its relay power denoted by $x^{(k)}\in\mathbf{A}=[0,P_U^M]$, where $P_U^M$ is the maximum UAV transmit power, and $\mathbf{A}$ is the action set. The UAV sends the message with power $x^{(k)}$ and relay cost denoted by $C_U$ to the backup BS, i.e., BS$_1$ in Fig. 1, if $x^{(k)}>0$. The backup BS relays the message from the UAV to the server via fibers. The server decodes the user message and measures the BER of the message from the user to the UAV denoted by $\rho_2^{(k)}$ and that of the message relayed by the UAV to the backup BS denoted by $\rho_3^{(k)}$, and sends $\rho_2^{(k)}$ and $\rho_3^{(k)}$ to the UAV via BS$_1$. For simplicity, we denote $\bm{\rho}^{(k)}=[\rho_i^{(k)}]_{1\leq i\leq3}$ as the BER vector.
\subsection{UAV-Ground Channel Model}
The UAV-ground links in cellular systems sometimes have shadow fading due to terrains and buildings and multi-path propagation due to the mountains, ground surface, and foliage \cite{Zeng2016Wireless}. The channel power gain of the user-BS$_0$ link denoted by $h_1^{(k)}$ is usually lower than the user-UAV link denoted by $h_2^{(k)}$ due to the LOS propagation of the UAV. Similarly, the channel power gain of the UAV-BS$_1$ link denoted by $h_5^{(k)}$ is much higher than that of the user-BS$_0$ link due to the LOS UAV-BS$_1$ channel. According to \cite{Xie2018User}, the UAV-ground channels follows a log-normal shadowing model with constant channel power gains within a time slot.

\subsection{Jamming Model}
A jammer is located close to the current serving BS of the user and sends jamming signals to prevent BS$_0$ from receiving messages from the user. As an advanced and most dangerous type of jammers, smart jammers can apply USRP to observe the BER of the user message, the user-BS$_0$/UAV link conditions, and the UAV relay power, and use RL to optimize the jamming policy with a goal of depleting the energy of the serving BS and the user. More specifically, a smart jammer close to BS$_0$ applies Q-learning to choose its jamming power $y^{(k)}\in[0,P_J^M]$, where $P_J^M$ is the maximum jamming power.
The channel power gain of the jammer-BS$_0$ link is denoted by $h_3^{(k)}$, and the jammer-UAV link is denoted by $h_4^{(k)}$. For simplicity, the channel power gain vector is denoted by $\mathbf{h}^{(k)}=[h_\varsigma^{(k)}]_{1\leq\varsigma\leq5}$, and the receiver noise power is denoted by $\sigma$.

\section{Deep Reinforcement Learning based UAV Relay Scheme}
In this article, we present a deep reinforcement learning based UAV relay scheme (DRLUR) to help cellular systems resist jamming attacks. By applying reinforcement learning, deep learning and transfer learning techniques, this scheme helps a UAV to achieve the optimal relay power without being aware of the network model and jamming model.

This scheme uses the weighted least squares algorithm in \cite{ChannelEstimation} to estimate the received jamming power denoted by $\hat{y}^{(k-1)}$, the jammer-UAV link condition $\hat{h}_4^{(k-1)}$, the user-BS$_0$ link condition $\hat{h}_1^{(k-1)}$, the user-UAV link condition $\hat{h}_2^{(k-1)}$, the jammer-BS$_0$ link condition $\hat{h}_3^{(k-1)}$, and the UAV-BS$_1$ link condition $\hat{h}_5^{(k-1)}$. For simplicity, we define the channel gain vector as $\hat{\bm{h}}^{(k-1)}=[\hat{h}_\varsigma^{(k-1)}]_{1\leq\varsigma\leq5}$. The relay power $x^{(k)}$ is chosen based on the state $\mathbf{s}^{(k)}$ that includes the BER vector $\bm{\rho}^{(k-1)}$ sent by the server, the channel gain vector $\hat{\bm{h}}^{(k-1)}$ and the estimated jamming power $\hat{y}^{(k-1)}$, with $\mathbf{s}^{(k)}=[\bm{\rho}^{(k-1)},\hat{\bm{h}}^{(k-1)},\hat{y}^{(k-1)}]$. As the next state observed by the UAV $\mathbf{s}^{(k+1)}$ is independent of the previous states and actions, for given current state and relay power. Therefore, the UAV relay process can be viewed as an MDP and thus the UAV can use reinforcement learning to optimize the relay power.
\begin{figure*}[!t]
\centering\includegraphics[height=2.8in]{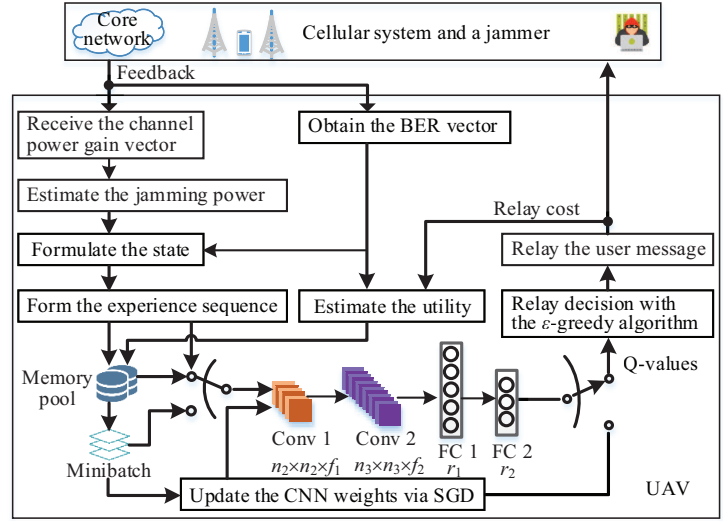}\\
\caption{Illustration of the deep RL based UAV relay scheme for cellular systems.}\label{fig:learning}
\end{figure*}

\begin{algorithm}[!t]\label{algorithm1}
\centering
\caption{Deep RL based UAV relay scheme}\label{DQN}
\begin{algorithmic}[1]
\STATE Initialize the action set $\mathbf{A}$, the learning parameters $E$, $M$, and $\mathcal{D}$
\STATE Initialize the CNN weights with the previous $\Gamma$ similar anti-jamming relay experiences
\FOR {$k=1, 2, ...$}
\STATE Receive the previous BER vector $\bm{\rho}^{(k-1)}$ from the server
\STATE Receive the channel gain vector from BS$_0$ and BS$_1$
\STATE Estimate the jamming power $\hat{y}^{(k-1)}$, the channel gain of the user-UAV link $\hat{h}_2^{(k-1)}$ and the jammer-UAV channel state $\hat{h}_4^{(k-1)}$
\STATE $\mathbf{s}^{(k)}=[\bm{\rho}^{(k-1)},\hat{\bm{h}}^{(k-1)},\hat{y}^{(k-1)}]$
\IF {$k\leq E$}
\STATE Choose $x^{(k)}$ randomly
\ELSE
\STATE Input $\varphi^{(k)}$ to the CNN
\STATE Update the Q-values with the CNN output
\STATE Choose $x^{(k)}$ with the $\varepsilon$-greedy algorithm
\ENDIF
\IF {$x^{(k)}>0$}
\STATE Relay the user message to BS$_1$ with a transmit power $x^{(k)}$
\ENDIF
\STATE Evaluate utility $u^{(k)}$
\STATE $\varphi^{(k+1)}=\{\mathbf{s}^{(k-E+2)},x^{(k-E+2)},\cdots,x^{(k)},\mathbf{s}^{(k+1)}\}$
\STATE $\mathcal{D}\leftarrow\mathcal{D}\cup\mathbf{e}^{(k)}$
\STATE Sample $M$ experiences from $\mathcal{D}$
\STATE Update the CNN weights $\theta^{(k)}$ via SGD
\ENDFOR
\end{algorithmic}
\end{algorithm}

This relay scheme uses CNN to compress the state space and applies a type of transfer learning called hotbooting as presented in \cite{Lu2018UAV} to initialize the CNN weights denoted by $\theta^{(k)}$ and learning parameters such as the learning rate with previous anti-jamming relay experiences. More specifically, the CNN weights are initialized via $\Gamma$ similar anti-jamming relay experiences each containing $K$ time slots. The experience sequence denoted by $\varphi^{(k)}$ consists of the previous $E$ states and the $E-1$ relay policies, which is then reshaped into an $n_1\times n_1$ matrix as the CNN.

As shown in Fig. \ref{fig:learning}, the CNN includes two convolutional (Conv) layers and two fully connected (FC) layers, in which Conv 1 has $f_1$ filters each with size $n_2\times n_2$ and stride 1, and Conv 2 has $f_2$ filters each with size $n_3\times n_3$ and stride 1. The two Conv layers use the rectified linear units (ReLUs) as the activation function, and the FC layers involve $r_1$ and $r_2$ ReLUs, respectively. The CNN outputs the Q-values of each relay policy for the experience sequence.

The UAV chooses its relay power $x^{(k)}$ that depends on the current state and the Q-values output by the CNN with the $\varepsilon$-greedy algorithm to avoid tracking in the local optimum at the beginning. More specifically, the $\varepsilon$-greedy algorithm is applied to choose the relay policy that maximizes the UAV utility with a high probability 1-$\varepsilon$, and select the other policies with a small probability. According to the selected relay power $x^{(k)}$, the UAV relays the message with a transmit power $x^{(k)}$ and relay cost $C_U$ to the backup BS if $x^{(k)}>0$, and keeps silent otherwise. The current UAV utility denoted by $u^{(k)}$ depends on the BER of the user message measured by the server and the relay cost.

The anti-jamming relay experience denoted by $\mathbf{e}^{(k)}=\{\varphi^{(k)},x^{(k)},u^{(k)},\varphi^{(k+1)}\}$ is saved in the memory pool $\mathcal{D}$, with $\mathcal{D}=\{\mathbf{e}^{(1)},\cdots,\mathbf{e}^{(k)}\}$. The UAV applies an experience replay technique to extract the anti-jamming relay experience in a memory pool at each time slot. Specifically, the UAV randomly samples $M$ experiences from the updated memory pool $\mathcal{D}$ in the experience replay. The stochastic gradient descent (SGD) algorithm is used to iteratively update the CNN weights $\theta^{(k)}$ similar to \cite{Xie2018User}, which minimizes the mean-squared error between the CNN output and the target optimal Q-value in Algorithm \ref{algorithm1}.


\section{Performance Analysis}
The performance of the UAV-aided cellular systems based on deep reinforcement learning can be evaluated in a dynamic UAV relay game, in which the jammer chooses its jamming power $y\in[0,P_J^M]$, and the UAV decides the relay power $x\in[0,P_U^M]$. For simplicity, we assume a quadrature phase-shift keying as the digital modulation in the cellular system with an additive white Gaussian noise. The BER of the user message denoted by $P_e^{(k)}$ at time slot $k$. The UAV utility depends on the BER of the user message received by BS$_0$, the BER of the weaker message signal received by UAV or BS$_1$, and the relay cost,
\begin{align}\label{utility}
&u^{(k)}=-\frac{1}{2}\mathrm{erfc}\Bigg(\max\Bigg(\sqrt{\frac{Ph_1^{(k)}}{\sigma+y^{(k)}h_3^{(k)}}},\notag \\
&\min\Bigg(\sqrt{\frac{Ph_2^{(k)}}{\sigma+y^{(k)}h_4^{(k)}}},\sqrt{\frac{x^{(k)}h_5^{(k)}}{\sigma}}\Bigg)\Bigg)\Bigg)-x^{(k)}C_U.
\end{align}

The performance lower bound of Algorithm 1 can be proved to be given by the Nash equilibrium (NE) in the UAV relay game. This scheme enables a UAV to optimize its relay power in the dynamic anti-jamming communication game without knowing the cellular network and jamming model. The time index $k$ is omitted in the superscript if no confusion occurs in this section.

The UAV applying Algorithm 1 to resist a weak jammer with a degraded channel to the backup BS does not relay the user message and the resulting BER is given by
\begin{align}\label{the1}
P_e=\frac{1}{2}\mathrm{erfc}\left(\sqrt{\frac{Ph_1}{\sigma}}\right).
\end{align}
Both the jammer and the UAV keep silent to save energy if the UAV relays the message to the backup BS with a degraded channel and the jammer attacks the system with weak jamming strategy. As shown in (\ref{the1}), the BER of the user message decreases with the user transmit power.

On the other hand, the UAV applying Algorithm 1 to resist a smart jammer with a better channel to the backup BS can achieve the optimal relay power and the resulting BER is given by
\begin{align}\label{the2}
P_e=&\frac{1}{2}\mathrm{erfc}\left(\max\left(\sqrt{\frac{Ph_1}{\sigma+h_3P_J^M}},\right.\right.\cr
&\left.\left.\min\left(\sqrt{\frac{Ph_2}{\sigma+h_4P_J^M}},\sqrt{\frac{P_U^Mh_5}{\sigma}}\right)\right)\right).
\end{align}
The UAV decides to relay the message to the backup BS with the maximum relay power if the channel condition between the UAV and the backup BS is good enough. In addition, the jammer with smart jamming strategy chooses the maximum jamming power to attack the UAV-aided cellular system.

The complexity of the deep RL based relay scheme as shown in Fig. \ref{fig:learning} depends on the CNN complexity. Similar to \cite{CVPR}, the CNN complexity depends on the number of input channels for the CNN, the number of the two Conv layer filters $f_1$ and $f_2$, and the spatial size of the output feature map of the Conv layer $i$. Thus, the computation complexity of the deep RL based UAV relay scheme is $O\left(f_1f_2n_3^2(n_1-n_2-n_3+2)^2\right)$ according to \cite{Xie2018User}.
\section{Simulation Results}
We performed simulations to evaluate the performance of Algorithm 1 in a cellular system against a jammer. This algorithm was implemented with python 3.5 on Dell Vostro 3900 with a deep learning platform PyTorch, with each simulation running 2.2 s. The simulations assumed the cellular channel model and network topology model similar to the 5G toolbox, with both the user-BS0 link and the jammer-BS0 link following the tapped delay line model, the frequency ranging from 0.5 GHz to 6 GHz and the bandwidth of 100 MHz. All the two BSs are uniformly and randomly distributed in the area. The user transmit power is $50$ mW and the UAV transmit power ranges between 0 and 150 mW. The user moves within the BS0 coverage following the random waypoint model. The smart jammer implemented on USRP with the goal to minimize the UAV utility with lower jamming cost measures the BER of the user message as the basis to choose the jamming power ranging from 0 to 80 mW.

The UAV relay scheme initializes CNN weights via $10$ anti-jamming relay experiences each containing $500$ time slots, considers previous $13$ state action pairs, with 16 pieces of sample experience, 20 filters in Conv 1, 40 filters in Conv 2, 6 filter size in Conv 1, 5 filter size in Conv 2, 1000 ReLUs in FC 1, and 31 ReLUs in FC 2 to maximize the UAV utility according to the transfer learning algorithm as presented in \cite{Lu2018UAV}.


\begin{figure}[!ttbp]
\centering
\subfigure[BER of the user messages]{
\label{fig:BER}
\includegraphics[height=2.3 in]{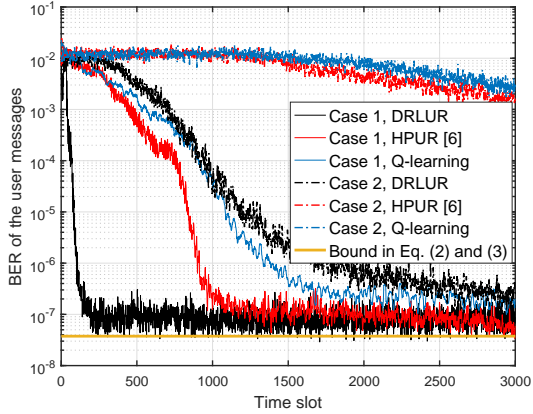}}\\
\subfigure[Energy consumption of the cellular system]{
\label{fig:energy}
\includegraphics[height=2.3 in]{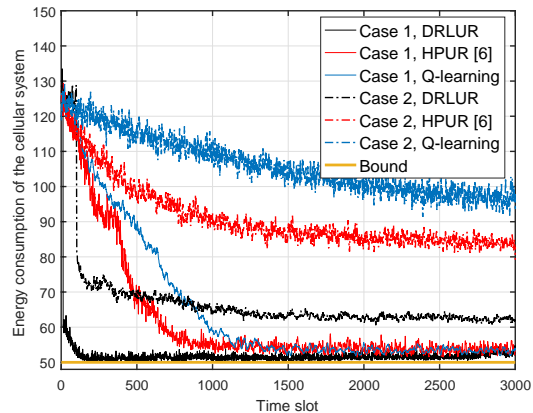}}\\
\caption{Performance of the UAV-aided cellular system against jamming, with Case 1 corresponding to the ideal learning environments and Case 2 for the UAV with state estimation delay of a time slot and error following Gaussian distribution $\mathcal{N}(0,1.5)$: a) BER of the user messages; b) Energy consumption of the cellular system.}\label{FIGG}
\end{figure}

The anti-jamming communication performance of Algorithm 1 is shown in Fig. \ref{FIGG}. As shown in Fig. \ref{fig:BER}, the proposed DRLUR has lower BER compared with HPUR \cite{Lu2018UAV} and Q-learning based relay algorithm. For example, DRLUR reduces the BER of the user messages by 46.6 percent and 99.7 percent at the 1000-th time slot, compared with HPUR and the Q-learning based relay algorithm, respectively. That's because DRLUR applies transfer learning technique to initialize the CNN parameters, and uses CNN to compress the high-dimensional state space to save the learning time of the dynamic UAV relay game. In addition, DRLUR can converge to the NE of the theoretical results, if the dynamic UAV relay game is long enough. Our proposed DRLUR is less sensitive to the state estimation error and delay compared with HPUR. For example, DRLUR saves the cellular energy consumption by 24.6 percent at time slot 1500 and reduces the BER by 92.8 at time slot 700 compared with HULR.

The proposed DRLUR saves the learning time and the cellular energy consumption compared with HPUR, as shown in Fig. \ref{fig:energy}. For example, DRLUR takes about 200 time slots (i.e., 0.22s) to optimize the relay power, which saves 84.6 percent of the time required by HPUR. In addition, DRLUR saves the energy consumption by 33.6 percent compared with HPUR. Our proposed scheme accelerates the learning speed of the dynamic UAV relay process, yielding a lower energy consumption compared with the benchmarks.

\section{Conclusions and Future Work}
In this article, we have proposed a deep RL based UAV relay scheme to optimize the relay power without knowing the jamming and the network model for cellular systems against jamming. We have provided the performance bound of the proposed relay scheme in terms of the BER and the energy consumption and evaluated its computation complexity. These analysis results have been verified via simulations, showing that this relay scheme can efficiently improve the jamming resistance of cellular systems. For instance, the proposed relay scheme reduces the BER by 44.6 percent and saves the cellular energy by 33.6 percent compared with HPUR as presented in \cite{Lu2018UAV}.

Several challenges have to be addressed to implement the deep RL based relay scheme in practical cellular systems:

The deep learning techniques such as DQN require a UAV to try all the policies in the learning process, and "bad" policies that sometimes cause network disasters for cellular systems. The dangerous UAV exploration can lead to the failure to send critical information for the users and to satisfy the quality of the service by the users. This issue can be addressed by transfer learning and data mining, which explore the anti-jamming communication defense experiences to reduce the random exploration and save the risks of trying dangerous UAV policies at the beginning of the learning process. Backup anti-jamming communication protocols have to be designed and incorporated with the deep RL based UAV relay scheme to provide reliable and secure cellular communications.

Another important issue for the deep RL based relay scheme is the state estimation error and delay of the UAV. The current analysis assumes that the UAV can accurately estimate the transmission performance of the cellular systems and evaluate the immediate utility in time, which does not hold for practical UAV-aided cellular systems. Therefore, our future work is focused on the design of the deep reinforcement learning based UAV schemes that are robust against the state estimation error and delay to resist jamming attacks for cellular systems.

\ifCLASSOPTIONcaptionsoff
  \newpage
\fi

\bibliography{reference}
\bibliographystyle{IEEEtr}

\balance
\end{document}